\documentclass[a4paper]{jpconf}
\usepackage{graphicx}
\usepackage{amsmath}

\newcommand{\bk}{{\bf k}}
\newcommand{\bq}{{\bf q}}

\newcommand{\bK}{{\bf K}}

\newcommand{\vF}{v_{\mathrm{F}}}

\newcommand{\modified}[1]{{\relax #1}}

\begin{document}
\title{Effect of uniaxial strain on plasmon excitations in graphene}

\author{F. M. D. Pellegrino$^{1,2}$, G. G. N. Angilella$^{1,2,3,4}$, R.
Pucci$^{1,2}$}

\address{$^1$ Dipartimento di Fisica e Astronomia, Universit\`a di Catania, Via
S. Sofia, 64, I-95123 Catania, Italy}
\address{$^2$ CNISM, UdR di Catania,  Via S. Sofia, 64, I-95123 Catania, Italy}
\address{$^3$ Scuola Superiore di Catania, Universit\`a di Catania, Via
Valdisavoia, 9, I-95123 Catania, Italy}
\address{$^4$ INFN, Sez. di Catania,  Via S. Sofia, 64, I-95123 Catania, Italy}

\ead{giuseppe.angilella@ct.infn.it}

\begin{abstract}
Uniaxial strain is known to modify significantly the electronic properties of
graphene, a carbon single layer of atomic width. Here, we study the effect of
applied strain on the composite excitations arising from the coupling of charge
carriers and plasmons in graphene, \emph{i.e.} the plasmarons. Specifically, we
predict that the plasmaron energy dispersion, which has been recently observed
experimentally in unstrained graphene, is shifted and broadened by applied
uniaxial strain. Thus, strain constitutes an additional parameter which may be
useful to tune graphene properties in plasmaronic devices.
\end{abstract}

Graphene is a monoatomic carbon layer with honeycomb structure. After having
been considered for decades as the ideal constituent of most compounds of carbon
in the $sp^2$ hybridization state, it has been recently obtained in the
laboratory \cite{Novoselov:05a}, thus kindling intense research activity both on
the experimental and on the theoretical side. Graphene is especially
characterized by a quasiparticle band structure consisting of two bands,
touching at the Fermi level in a linear, cone-like fashion at the so-called
Dirac points $\pm\bK$, and a linearly vanishing density of states (DOS) at the
Fermi level \cite{CastroNeto:08,Abergel:10}. These peculiar electronic
properties, along with the reduced dimensionality, have remarkable effects on
the electromagnetic properties of graphene. These include, \emph{e.g.,} the
reflectivity \cite{Nair:08}, the optical conductivity
\cite{Kuzmenko:08,Wang:08,Mak:08,Stauber:08a,Pellegrino:09b}, the plasmon
dispersion relation \cite{Hwang:07a,Polini:09,Pellegrino:10a,Pellegrino:10c}, as
well as a newly predicted transverse electromagnetic mode \cite{Mikhailov:07},
which is characteristic of a 2D system with a double band structure, such as
graphene.

The composite elementary excitations arising from the coupling of charge
carriers and plasmons, the so-called plasmarons, have been considered in a
general context earlier on by Lundqvist \cite{Lundqvist:67,Lundqvist:67a}.
Recently, plasmarons have been experimentally observed in graphene by means of
angular resolved photoemission spectroscopy (ARPES) \cite{Bostwick:10}, and
their dispersion relation described theoretically within an improved version of
the random phase approximation ($G_0 W$-RPA) \cite{Polini:08}.

In $n$ doped graphene, a plasmaron mode with momentum $\bk$ results from the
relatively strong coupling of a quasihole with momentum $\bk+\bq$, and a plasmon
with momentum $-\bq$, the quasihole-plasmon coupling being stronger when the two
excitations have the same group velocity \cite{Polini:08}. At $\bk=0$, the
plasmaron relative momentum modulus turns out to be
\begin{equation}
q= \frac{e^2}{8\pi\epsilon}\frac{\mu}{(\hbar\vF)^2} ,
\label{eq:q0}
\end{equation}
where $\mu$ is the chemical potential, $\vF$ the Fermi velocity,
$\epsilon=\epsilon_0 \epsilon_r$ the dielectric constant. Therefore, the
plasmaron binding energy with respect to the Fermi energy can be estimated, in
first approximation, as the sum of the energies of the bare quasihole and
plasmon, both having momentum modulus $q$, \emph{viz.}
\begin{equation}
E_P = -\mu -\alpha \frac{c}{\vF} \frac{\mu}{2\epsilon_r} ,
\label{eq:EP0}
\end{equation}
where $\alpha$ is the fine structure constant. \modified{In the realistic case
of graphene on a SiO$_2$ substrate, Eq.~(\ref{eq:EP0}) yields $E_P
\simeq - 1.25 \mu$. A more accurate estimate, including the contribution of the
quasihole-plasmon interaction at the $G_0 W$-RPA level \cite{Polini:08}, yields
$E_P \simeq -1.3\mu$, in better agreement with the experimental results
\cite{Bostwick:10}.}

We now consider the effect of uniaxial strain on the graphene sheet. It has been
shown that this amounts to a shift of the Dirac points in reciprocal space, and
to an anisotropic deformation of the Dirac cone centred at those points
\cite{Pellegrino:11}, with elliptic sections at constant energy. Such a
strain-induced angular dependence may also be nonuniform, as in the case of
coordinate-dependent strain \cite{Pellegrino:11c}. The effect of applied
uniaxial strain on the plasmon dispersion relation of graphene has been studied
in Refs.~\cite{Pellegrino:10a,Pellegrino:10c}. Besides, is has been shown
\cite{Pellegrino:11} that strain may significantly modify the dispersion
relation of a transverse plasmon mode, which has been recently predicted to
occur in graphene \cite{Mikhailov:07}. Therefore, it can be expected, on general
grounds, that strain affects the energy dispersion of the plasmaronic modes.

Denoting by $\varepsilon$ the modulus of applied strain, $\theta$ its direction
(with $\theta=0$ and $\theta=\pi/6$ referring to strain along the zig~zag and
armchair directions, respectively), and $\nu$ the Poisson's ratio of graphene
($\nu=0.14$ \cite{Farjam:09}, to be compared with the known experimental value
$\nu=0.165$ for graphite \cite{Blakslee:70}, and with $\nu=-1$, corresponding to
the hydrostatic limit), following Ref.~\cite{Pellegrino:11}, the quasiparticle
dispersion relation, to linear order in $\varepsilon$, reads
\begin{equation}
\epsilon_\bq = \pm \hbar \vF q [\left(1-\kappa(1-\nu)\varepsilon\right)
-\kappa(1+\nu)\varepsilon \cos(2\theta+2\varphi)],
\label{eq:epsq}
\end{equation}
where the $+$ ($-$) sign refers to the conduction (valence) band, and while the
plasmon dispersion relation under strain becomes
\begin{equation}
\hbar\omega_{\mathrm{pl}} = \sqrt{\frac{e^2}{2\pi\epsilon}\mu}
\left[ 1 - \kappa(1+\nu)\varepsilon \cos(2\theta-2\varphi)\right] \sqrt{q} .
\label{eq:omegapl}
\end{equation}
Deriving the corresponding group velocities from the above Eqs.~(\ref{eq:epsq})
and (\ref{eq:omegapl}), Eqs.~(\ref{eq:q0}) and (\ref{eq:EP0}) for the plasmaron
momentum and energy, respectively, get modified into
\begin{subequations}
\begin{eqnarray}
q &=& [1+2\kappa(1-\nu)\varepsilon]
\frac{e^2}{8\pi\epsilon}\frac{\mu}{(\hbar\vF)^2} ,\\
\label{eq:EP}
E_P (\varphi) &=& -\mu -\alpha \frac{c}{\vF} \frac{\mu}{2\epsilon_r}
[1+\kappa(1-\nu)\varepsilon -\kappa(1+\nu)\varepsilon\cos(2\theta-2\varphi)] ,
\end{eqnarray}
\end{subequations}
to linear order in the strain modulus $\varepsilon$. Here, $\kappa =
(a/2t)|\partial t/\partial a| -\frac{1}{2} \approx 1.1$ is related to the
logarithmic derivative of the nearest-neighbor hopping $t$ at $\varepsilon=0$,
$a$ is the carbon--carbon distance, and $\theta$ is the direction of the
stress. 

Eq.~(\ref{eq:EP}) shows that, in the presence of applied uniaxial strain, the
plasmaronic energy at $\bk=0$ acquires an explicit dependence on the angle
$\varphi$ of the quasihole momentum $\bq$. This is due to the anisotropy of both
the electronic and the plasmon spectrum. Correspondingly, the plasmaron energy
is characterized by a central value
\begin{equation}
E_P^{\mathrm{c}} = -\mu -\alpha \frac{c}{\vF} \frac{\mu}{2\epsilon_r}
[1+\kappa(1-\nu)\varepsilon] ,
\end{equation}
and a strain-induced energy spread
\begin{equation}
\Delta E_P= \alpha\frac{c}{v_F}\frac{\mu}{\epsilon_r} \kappa(1+\nu)\varepsilon.
\end{equation}
\modified{Considering again the realistic case of graphene on a
SiO$_2$ substrate, one can estimate the central plasmaron energy in the
unstrained case as $E_P^{\mathrm{c}} (\varepsilon=0) = -1.25\mu = -125$~meV, for
$\mu=100meV$, with zero energy spread. Correspondingly, in the case of an
applied strain $\varepsilon=10$~\%, one finds a central plasmaron energy of
$E_P^{\mathrm{c}} (\varepsilon=10~\%) = -127.4$~meV, with an energy spread
$\Delta E_P (\varepsilon=10~\%) = 6.27$~meV.}

In conclusion, the effect of applied uniaxial strain on graphene is therefore
that of shifting and broadening the plasmaron energy, proportionally to the
strain modulus.  Therefore, by suitably applying uniaxial strain, one gains
further control on the energy of the plasmaronic excitation, besides the
possibility of tuning the relative dielectric constant $\epsilon_r$
\cite{Walter:11}. This may be instrumental for the realization of `plasmaronic'
devices.

\section*{References}

\bibliographystyle{jpconf}
\bibliography{a,b,c,d,e,f,g,h,i,j,k,l,m,n,o,p,q,r,s,t,u,v,w,x,y,z,zzproceedings,Angilella}

\end{document}